\documentclass[apjl]{emulateapj}
\usepackage{amssymb}
\usepackage{epsfig}
\usepackage{url}
\usepackage{enumerate}

\newcommand{\etal}{et~al.~}

\begin{document}


\title{HST-COS Spectroscopy of the Cooling Flow in Abell\,1795 -- Evidence for Inefficient Star Formation in Condensing Intracluster Gas}


\author{Michael McDonald$^{1*\dagger}$, Joel C.\ Roediger$^2$, Sylvain Veilleux$^{3,4}$, Steven Ehlert$^{1}$}
\altaffiltext{1}{Kavli Institute for Astrophysics and Space Research, MIT, Cambridge, MA 02139, USA}
\altaffiltext{2}{Department of Astronomy, University of California Santa Cruz, CA 95064, USA}
\altaffiltext{3}{Department of Astronomy, University of Maryland, College Park, MD 20742, USA}
\altaffiltext{4}{Joint Space-Science Institute, University of Maryland, College Park, MD 20742, USA}

\altaffiltext{*}{Email: mcdonald@space.mit.edu}
\altaffiltext{$\dagger$}{Hubble Fellow}


\begin{abstract}
We present far-UV spectroscopy from the \emph{Cosmic Origins Spectrograph} on the \emph{Hubble Space Telescope} of a cool, star-forming filament in the core of Abell~1795. These data, which span 1025\AA~$<$~$\lambda_{rest}$~$<$ 1700\AA, allow for the simultaneous modeling of the young stellar populations and the intermediate-temperature (10$^{5.5}$\,K) gas in this filament, which is far removed ($\sim$30\,kpc) from the direct influence of the central AGN. Using a combination of UV absorption line indices and stellar population synthesis modeling, we find evidence for ongoing star formation, with the youngest stars having ages of $7.5^{+2.5}_{-2.0}$\,Myr and metallicities of 0.4$^{+0.2}_{-0.1}$\,Z$_{\odot}$. The latter is consistent with the local metallicity of the intracluster medium. We detect the O\,\textsc{vi}\,$\lambda$1038 line, measuring a flux of $f_{\scriptsize \textrm{O}\,\tiny\textrm{VI},1038}$ = 4.0 $\pm$ 0.9 $\times$ 10$^{-17}$ erg s$^{-1}$ cm$^{-2}$. The O\,\textsc{vi}\,$\lambda$1032 line is redshifted such that it is coincident with a strong Galactic H$_2$ absorption feature, and is not detected. The measured O\,\textsc{vi}\,$\lambda$1038 flux corresponds to a cooling rate of 0.85 $\pm$ 0.2 (stat) $\pm$ 0.15 (sys) M$_{\odot}$ yr$^{-1}$ at $\sim10^{5.5}$\,K, assuming that the cooling proceeds isochorically, which is consistent with the classical X-ray luminosity-derived cooling rate in the same region. We measure a star formation rate of 0.11 $\pm$ 0.02 M$_{\odot}$ yr$^{-1}$ from the UV continuum, suggesting that star formation is proceeding at $13^{+3}_{-2}$\% efficiency in this filament. We propose that this inefficient star formation represents a significant contribution to the larger-scale cooling flow problem.
\end{abstract}


\keywords{ galaxies: clusters: individual (Abell~1795); galaxies: clusters: intracluster medium; galaxies: star formation;  galaxies: stellar content}


\section{Introduction}

In the cores of galaxy clusters, the intracluster medium (ICM) can reach high enough density and low enough temperature that the inferred cooling time is much shorter than the age of the Universe. In these so-called ``cooling flow clusters'', simple models predict that $\sim$100--1000 M$_{\odot}$ yr$^{-1}$ of cool gas should condense out of the hot intracluster plasma and fuel star formation in the central cluster galaxy \citep[for a review, see][]{fabian94}. However, with few notable exceptions \citep{mcnamara06,mcdonald12c}, we do not observe such vigorous starbursts at the centers of galaxy clusters: the typical star formation rate in the core of a cooling flow cluster is only $\sim$1--10 M$_{\odot}$ yr$^{-1}$ \citep[e.g.,][]{hicks05,odea08,mcdonald11b}. This low level of star formation is most likely being fueled by local thermomodynamic instabilities in the ICM \citep[e.g.,][]{mccourt12}, with the remaining $\gtrsim90\%$ of the energy lost from cooling being offset by radio-mode feedback from the central AGN \citep[e.g.,][]{churazov01, rafferty06, rafferty08,fabian12,mcnamara12}.

While most cooling flow clusters show evidence of such ``reduced cooling flows'' in the form of ongoing star formation \citep[e.g.,][]{johnstone87,mcnamara89,allen95,odea08,hicks10,mcdonald11b} and cold molecular gas \citep[e.g.,][]{edge01,edge02,salome03,hatch05,salome11,mcdonald12b}, there is still very little evidence for gas between 10$^4$\,K and 10$^7$\,K, which would directly link the hot and cool phases. 
High resolution X-ray spectroscopy of individual clusters has, thus far, only been able to put upper limits on the amount of $\sim$10$^6$\,K gas in the cores of galaxy clusters \citep[e.g.,][]{peterson03,peterson06,sanders10}. 
Recently, \cite{sanders11b} performed a stacking analysis of X-ray grating spectra, yielding a detection of O\,\textsc{vii} at a level $\sim$4--8 times lower than the expectation from simple cooling flow models. 
Using the FUSE satellite, \cite{oegerle01} and \cite{bregman01,bregman06} found evidence for O\,\textsc{vi} emission in the far-UV, probing gas at $\sim10^{5.5}$\,K, in the cores of several galaxy clusters, inferring cooling rates well below the cooling flow expectation. These observations were nearly always centered on the nucleus of the central cluster galaxy, where AGN feedback could in fact be heating the gas, causing excess O\,\textsc{vi} emission.

\begin{figure*}[htb]
\centering
\includegraphics[width=0.99\textwidth]{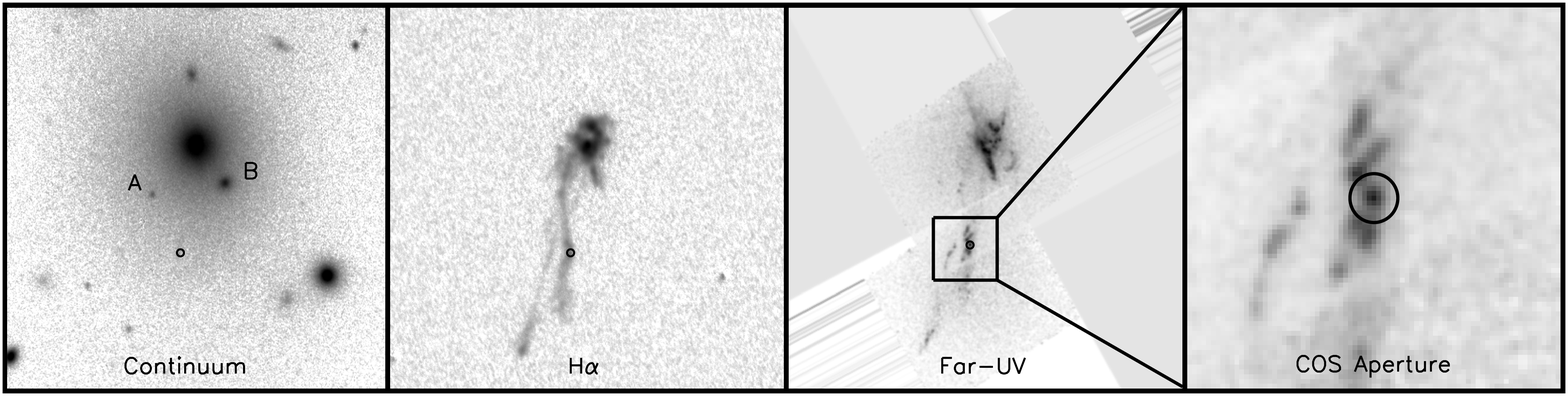}
\caption{\emph{Left:} Continuum ($\sim7000$\AA) image of the central galaxy in Abell~1795. \emph{Left-Center:} H$\alpha$ image showing the twin filaments extending $\sim$50\,kpc to the south of the central galaxy \citep{mcdonald09}. \emph{Right-Center:} Far-UV image from HST ACS/SBC \citep{mcdonald09} showing young stellar populations along the H$\alpha$ filaments. \emph{Right:} Zoom-in on the far-UV image, showing the location of the 2.5$^{\prime\prime}$ COS aperture. This aperture, which is shown in all four panels, is centered on the brightest UV clump in the western filament. In the left-most panel, the positions of the two closest galaxies are denoted by A and B. These galaxies are 16 and 21 kpc away in projection from the COS pointing, respectively, implying that any gas/stars in the filaments that originated in the satellite galaxies would have been stripped from these galaxies $>$26 Myr ago (assuming $v = 800$ km s$^{-1}$).
}
\label{fig:image}
\end{figure*}

Here we present new far-UV (FUV) spectroscopy (\S2) of Abell~1795 (A1795), a nearby, strongly-cooling galaxy cluster \citep[see e.g.,][Ehlert \etal 2014]{fabian01,ettori02}. These spectra were obtained along the southwestern filament, which is observed to be cooling rapidly in the X-ray \citep[][Ehlert \etal 2014]{crawford05,mcdonald10}, contains warm ($10^4$\,K) ionized \citep{cowie83,crawford05,mcdonald09} and cold molecular gas \citep{salome04,mcdonald12b}, and is rapidly forming stars\citep[$\sim$1 M$_{\odot}$ yr$^{-1}$;][]{mcdonald09}. With deep FUV spectroscopy, we can estimate the age and metallicity of these newly-formed stars (\S3), allowing us to determine if they are forming in situ or have been tidally stripped, while also measuring the O\,\textsc{vi} emission line flux (\S4) far from the influence of the central AGN ($\sim$30 kpc). This approach will allow us to link the young stars to the cooling X-ray gas, if that is indeed their origin (\S5). We will finish, in \S6, with a discussion of the current state of the cooling flow problem, and how these new data can advance our understanding.

Throughout this paper we assume H$_0$ = 70 km s$^{-1}$ Mpc$^{-1}$, $\Omega_M=0.27$, and $\Omega_{\Lambda}=0.73$.

\section{Data}
\setcounter{footnote}{0}
FUV spectroscopy for this program was acquired using the \emph{Cosmic Origins Spectrograph} (COS) on the \emph{Hubble Space Telescope} using the G140L grating with $\lambda_{center}=1280$\AA, which yields a spectral coverage of 1080--1900\AA. The 2.5$^{\prime\prime}$ aperture was centered at ($\alpha$, $\delta$) = 207$^{\circ}\hspace{-0.04in}.2199$, +26$^{\circ}$\hspace{-0.04in}.5873, which corresponds to the peak of both the FUV and H$\alpha$ emission along the filament (see Figure \ref{fig:image}).

COS spectroscopy is simultaneously obtained in ``blue'' and ``red'' channels, with respective wavelengths spanning $\sim$1080--1200\AA\ and $\sim$1250--1800\AA\ for our setup. The gap between 1200--1250\AA\ spans the geocoronal Ly$\alpha$ line (1216\AA). We observe one other strong geocoronal line due to O\,\textsc{i} at $\sim$1302.2--1306\AA\footnote{\url{http://www.stsci.edu/hst/cos/calibration/airglow\_table.html}}, which is redward of redshifted Ly$\beta$ ($\lambda_{Ly\beta} = 1290.8$\AA). We determine the redshift of the spectrum using a joint fit to the Ly$\beta$ emission line and two absorption features in the red channel (Si\,\textsc{iv}$\lambda$1394 and C\,\textsc{iv}$\lambda$1548). This fit yields $z=0.0619 \pm 0.0005$, which is consistent with our optical redshift of $z=0.0618$ from \cite{mcdonald12a} at the same position along the filament.

\section{FUV Continuum: Young Stellar Populations}
\subsection{UV Absorption Indices}

To constrain the age and metallicity of the stellar population responsible for the observed FUV continuum in A1795, we use predicted UV absorption line strengths from \cite{maraston09} [hereafter M09], which are cast in terms of the \emph{International Ultraviolet Explorer} (IUE) index system established by \cite{fanelli92}.  
The use of an index-based approach helps reduce the uncertainty in our results due to reddening effects, which are most severe at UV wavelengths.  Since the M09 models do not consider contributions of other hot stellar phases (e.g. blue horizontal branch), we implicitly assume that the observed FUV emission is entirely due to young ($<$ 1 Gyr) stars.  

\begin{deluxetable*}{c c c}[h!]
\centering
\tablecaption{Stellar Age and Metallicity from UV Absorption Line Indices}
\tablewidth{12.5 cm}
\tablehead{
\colhead{Indices} &
\colhead{~~A [Myr]~~} &
\colhead{log$_{10}$(Z) [Z$_{\odot}$]}  
}
\startdata

BL1302, Si\,\textsc{iv}, BL1425, C\,\textsc{iv}a, C\,\textsc{iv}, C\,\textsc{iv}e, BL1617, BL1664 & 7.5$^{+2.5}_{-2.0}$ & $-0.4^{+0.2}_{-0.1}$ \\
\\
\phantom{BL1302,} Si\,\textsc{iv}, BL1425, C\,\textsc{iv}a, C\,\textsc{iv}, C\,\textsc{iv}e, BL1617, BL1664 & 10.0$^{+10.0}_{-4.5}$ & -0.2$^{+0.2}_{-0.2}$ \\
BL1302, \phantom{Si\,\textsc{iv},} BL1425, C\,\textsc{iv}a, C\,\textsc{iv}, C\,\textsc{iv}e, BL1617, BL1664 & 7.5$^{+2.5}_{-2.0}$ & -0.4$^{+0.2}_{-0.3}$ \\
BL1302, Si\,\textsc{iv}, \phantom{BL1425,} C\,\textsc{iv}a, C\,\textsc{iv}, C\,\textsc{iv}e, BL1617, BL1664 & 7.5$^{+2.0}_{-4.0}$ & -0.4$^{+0.2}_{-0.1}$ \\
BL1302, Si\,\textsc{iv}, BL1425, \phantom{C\,\textsc{iv}a,} C\,\textsc{iv}, C\,\textsc{iv}e, BL1617, BL1664 & 9.5$^{+0.5}_{-2.0}$ & -0.5$^{+0.1}_{-0.1}$ \\
BL1302, Si\,\textsc{iv}, BL1425, C\,\textsc{iv}a, \phantom{C\,\textsc{iv},} C\,\textsc{iv}e, BL1617, BL1664 & 3.0$^{+4.0}_{-1.0}$ & -0.1$^{+0.5}_{-0.3}$ \\
BL1302, Si\,\textsc{iv}, BL1425, C\,\textsc{iv}a, C\,\textsc{iv}, \phantom{C\,\textsc{iv}e,} BL1617, BL1664 & 6.5$^{+3.5}_{-3.0}$ & -0.4$^{+0.2}_{-0.1}$ \\
BL1302, Si\,\textsc{iv}, BL1425, C\,\textsc{iv}a, C\,\textsc{iv}, C\,\textsc{iv}e, \phantom{BL1617,} BL1664 & 7.5$^{+2.5}_{-2.0}$ & -0.4$^{+0.1}_{-0.1}$ \\
BL1302, Si\,\textsc{iv}, BL1425, C\,\textsc{iv}a, C\,\textsc{iv}, C\,\textsc{iv}e, BL1617\phantom{, BL1664} & 7.5$^{+2.5}_{-2.0}$ & -0.4$^{+0.2}_{-0.1}$
\enddata
\enddata
\tablecomments{All uncertainties are 1$\sigma$. See \cite{maraston09} for a discussion of far-UV absorption indices.
}
\label{table:indices}
\end{deluxetable*}

\begin{figure*}[htb]
\centering
\includegraphics[width=0.99\textwidth]{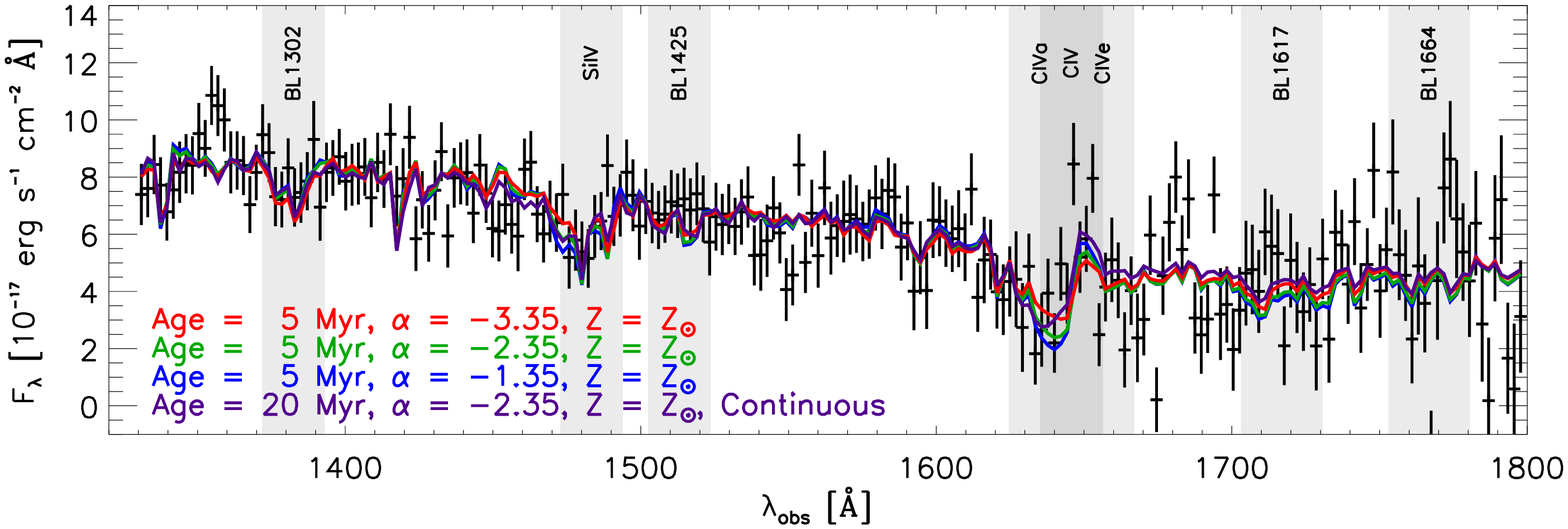}\\
\caption{Far-UV spectrum from the red channel of our HST-COS observation. This spectrum has been binned in wavelength by a factor of 27 (2.2\AA) in order to improve signal-to-noise. Vertical grey bands correspond to spectral indices as defined by \cite{fanelli92}. Overplotted are various models generated with Starburst99 \citep{leitherer99} using the latest stellar tracks from \cite{ekstrom12} and \cite{georgy13}, demonstrating the overall quality of these fits to the data. }
\label{fig:redcont}
\end{figure*}

The M09 models come in two flavors: ones based on empirical fitting functions 
to IUE spectra of Milky Way and Magellanic Cloud (MC) stars or others based on 
theoretical Kurucz spectra.  We use the latter here given that they reproduce 
the ages of MC globular clusters (from color-magnitude diagrams) to within a 
mean residual of 0.02 $\pm$ 0.32 dex (see Appendix C of M09).  These models 
span a semi-regular grid of ages from 1 Myr-1 Gyr and total metallicities from 
-1.00 to +0.35 dex with respect to solar\footnote{The spacing of the age grid changes from 0.5 Myr, 
to 5 Myr, to 50 Myr in the intervals 1 Myr-10 Myr, 10 Myr-0.1 Gyr, and 0.1-1.0 
Gyr, respectively.  On the other hand, the models are spaced at 0.1 dex 
intervals in terms of metallicity, except the +0.35 dex models, which are 
offset by 0.15 dex from the +0.20 dex models.}.  We allow age and metallicity 
to vary simultaneously in our fits, which follow a maximum likelihood 
approach.  Statistical errors in the best-fit quantities are estimated via 
Monte Carlo simulations of the measured index errors.  Although M09 recommend 
fitting to all IUE indices in the 1000-2000\AA\ range at once with their 
theoretical models, we have performed fits to a variety of index combinations, specifically a blue set (BL1302, Si\,\textsc{iv}, BL1425) and a red set (C\,\textsc{iv}a, C\,\textsc{iv}, C\,\textsc{iv}e, BL1617, BL1664), in order to test the robustness of our results. These results are summarized in Table \ref{table:indices}. Overall, we find consistently low ages (7.5$^{+2.5}_{-2.0}$ Myr) and metallicities (0.4$^{+0.2}_{-0.1}$ Z$_{\odot}$) regardless of which combination of indices we employ.

\subsection{Full Spectrum Synthesis}

\begin{figure}[htb]
\centering
\includegraphics[width=0.48\textwidth]{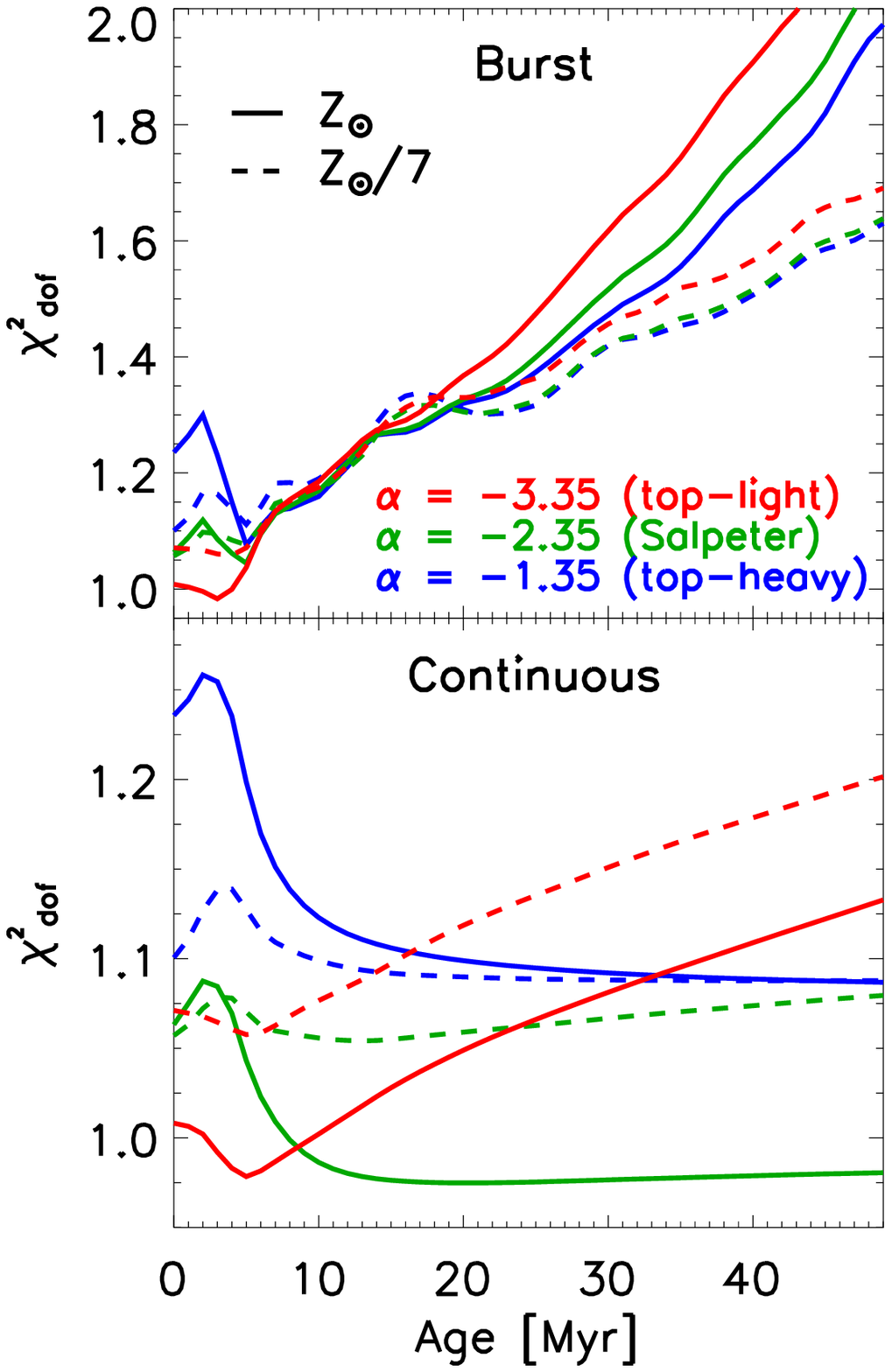}
\caption{Goodness-of-fit ($\chi^2_{dof}$) as a function of age for instantaneous (upper panel) and continuous (lower panel) star formation. For all choices of IMF and metallicity, the data are well-represented by either a young ($\lesssim$10\,Myr) stellar population or ongoing star formation.}
\label{fig:chis}
\end{figure}

A complementary method of constraining the age and metallicity of the UV-emitting stellar population is to model the full spectrum using synthetic stellar population models. For this, we opt to use the latest version of Starburst99 \citep[v7.0.0;][]{leitherer99}\footnote{\url{http://www.stsci.edu/science/starburst99/docs/default.htm}}, which is, to our knowledge, the only publicly-available stellar population synthesis (SPS) code with spectral resolution better than 1\AA\ over 1000\AA\ $< \lambda < 2000$\AA. We restrict this analysis to $>$1100\AA\ -- at bluer wavelengths, Starburst99 uses empirical stellar spectra which provide a qualitatively-poor fit to the data (see \S4). We use the latest Geneva tracks, which are only available for solar \citep{ekstrom12} and 0.14Z$_{\odot}$ \citep{georgy13} metallicities. We explore three different initial mass functions: Salpeter \citep[$\alpha=-2.35$;]{salpeter55}, with top-heavy ($\alpha=-1.35$) and top-light ($\alpha=-3.35$) variants. For each choice of metallicity and IMF, we generate synthetic spectra assuming either a burst or continuous star formation, over timescales of 50\,Myr. These model spectra are fit to the data, allowing the normalization, redshift, and reddening to vary, with a lower limit of the Galactic value imposed on the reddening \citep{schlegel98}.

Figures \ref{fig:redcont} and \ref{fig:chis} show the results of this exercise. In Figure \ref{fig:redcont} we compare several best-fitting models to the data, demonstrating the overall quality of these fits. These synthetic spectra are able to adequately fit the various absorption features discussed in \S3.1. The goodness-of-fit ($\chi^2_{dof}$) is shown in Figure \ref{fig:chis} as a function of age, metallicity, and IMF, for both instantaneous and continuous star formation. 
All starburst models favor a relatively young population, with the majority showing minima at ages of 5 Myr. The continuous star formation models prefer either a Salpeter IMF, with ages $>$10~Myr or a younger top-light stellar population. As is also the case for the burst-like models, the distinguishing power between IMFs comes from the C\,\textsc{iv} feature (see Fig.\ \ref{fig:redcont}), which is likely contaminated by emission. Qualitatively, all three choices of IMF perform equally well in fitting the spectrum over the range $1220 < \lambda_{rest} < 1700$\AA.

This analysis corroborates the results from FUV line indices (\S3.1), demonstrating that the best-fitting stellar population is one that is either currently forming stars, or ceased very recently ($\lesssim$10 Myr ago).

\section{FUV Emission Lines: Probing 10$^{5.5}$K Gas with O\,\textsc{vi}}

\begin{figure}[htb]
\centering
\includegraphics[width=0.49\textwidth]{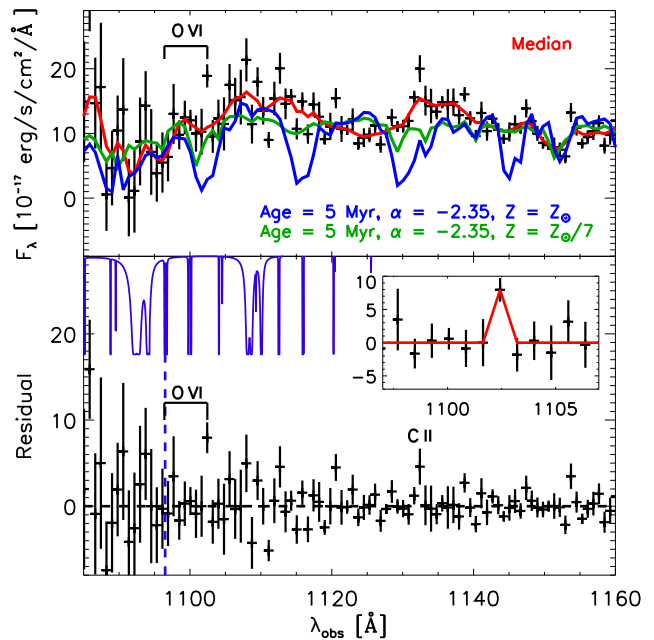}
\caption{\emph{Upper panel:} Far-UV spectrum from the blue channel of our HST-COS observation. Here, we show the binned (0.8\AA) data along with stellar population synthesis models from Starburst99 \citep{leitherer99} and a median continuum spectrum computed with a 2.4\AA\ smoothing window. At these wavelengths Starburst99 only provides empirical models, which yield poor fits to the data (note that the red channels are well-fit by a solar-metallicity model; Fig.\ \ref{fig:redcont}). \emph{Lower panel:} Residual spectrum, with the median continuum level subtracted. The expected location of the O\,\textsc{vi}\,$\lambda\lambda$1032,1038 doublet is highlighted. In purple, we plot the spectrum of Galactic H$_2$ absorption from \cite{mccandliss03}, which shows that the bluer O\,\textsc{vi}\,$\lambda$1032 line is most likely being absorbed by molecular gas in our galaxy. The correspondence between a saturated absorption line and the O\,\textsc{vi}$\lambda$1032 line is highlighted by a vertical dashed line. We detect the redder, unabsorbed line at $>$4$\sigma$ significance, as shown in the inset, with a velocity width of $\lesssim$0.7\AA\ ($\lesssim$90 km s$^{-1}$).
}
\label{fig:bluecont}
\end{figure}

In Figure \ref{fig:bluecont} we show the blue side ($<$1200\AA) of the spectrum. These data are significantly noisier than the red channels, and suffer from airglow emission and Galactic H$_2$ absorption to a higher degree. Despite this, there is evidence for emission at redshifted O\,\textsc{vi}\,$\lambda1038$ in the binned spectrum (upper panel of Fig.\ \ref{fig:bluecont}).  The spectrum of Galactic H$_2$ absorption from \cite{mccandliss03} shows strong absorption at the same wavelength as redshifted O\,\textsc{vi}\,$\lambda$1032, which likely explains the lack of emission from the bluer line in this doublet, which should have a factor of 2 higher flux. At these wavelengths, Starburst99 relies on empirical stellar spectra (see \S3) which provide a relatively poor fit to the data, so we opt to model the continuum spectrum by performing a median smoothing over a 2.4\AA\ ($\sim$650 km s$^{-1}$) window, which should be significantly wider than any emission lines. The residual spectrum with this model subtracted is shown in the lower panel of Figure \ref{fig:bluecont}. 

Assuming a linewidth of 0.43\AA\ ($\sigma = 50$ km s$^{-1}$), based on the H$\alpha$ velocity dispersion \citep{mcdonald12a}, we measure a flux of $f_{O\textsc{vi}} = 4.0 \pm 0.9 \times 10^{-17}$ erg s$^{-1}$ cm$^{-2}$ in the redder line of the O\,\textsc{vi} doublet ($>$4$\sigma$ detection). While the statistical significance of this line is low, we note that it is the most statistically significant deviation over the entire blue spectrum. The fact that this deviation is at the wavelength that we expect to find redshifted O\,\textsc{vi} further strengthens the significance of the detection. The width of this line is $\lesssim$0.7\AA\, or $\lesssim$90 km s$^{-1}$.







\section{Interpretation: Ongoing Star Formation in Condensing Filaments of Intracluster Gas}
Given the high ionization energy of the O\,\textsc{vi} transition (138.1 eV), it not likely to be due to the same process(es) responsible for the low-ionization H$\alpha$ emission (Fig.\ \ref{fig:image}). The original detection of O\,\textsc{vi} in the center of Abell~1795 by \cite{bregman06} was taken as evidence for cooling of the ICM. Alternatively, \cite{sparks12} recently invoked an evaporation scenario to explain the filamentary coronal emission in the core of the Virgo cluster. In the latter scenario, the hypothesis is that cool gas has been stripped from nearby gas-rich galaxies and is being heated via conduction by the hot ICM. The lack of stars in the filaments surrounding M87 was offered as further evidence of this scenario. 

The detection of young ($\lesssim$10 Myr) stars and cold molecular gas \citep{mcdonald12b} in the filaments of Abell~1795 favors the scenario in which gas is condensing and stars are forming in situ. If, instead, the cool gas originated in a nearby galaxy, it would have been stripped \emph{at least} 26 Myr ago, given the distance to the nearest satellite galaxy (21~kpc; Fig.\ \ref{fig:image}) and the typical galaxy velocities in the core of Abell~1795 ($\sim$800 km s$^{-1}$).
Assuming the stars are forming in situ, it seems unlikely that the filamentary gas is simultaneously condensing (into stars) and evaporating (causing coronal emission) -- the simplest explanation is that the hot intracluster gas is cooling through the O\,\textsc{vi} transition and into molecular gas, before ultimately forming stars. The agreement between the measured metallicity of the stars (0.4$^{+0.2}_{-0.1}$\,Z$_{\odot}$; Table 1) and the cooling ICM at the same position (0.5$^{+0.3}_{-0.2}$\,Z$_{\odot}$; Ehlert \etal 2014) further supports this scenario.

If we assume that the O\,\textsc{vi}\,$\lambda$1038 emission is due to cooling intracluster gas, we can estimate the cooling rate following \cite{edgar86} and \cite{voit94}. The inferred cooling rate can vary by a factor of $\sim$1.6 depending on if the cooling proceeds isobarically or isochorically. Based on very deep X-ray data of the cooling filament (Ehlert \etal 2014), we estimate the sound-crossing time at the location of the COS aperture to be $\sim$2-5~Myr and the cooling time to be $\sim$2~Myr (assuming $n_e \sim1$ $cm^{-3}$). Thus, it is unclear exactly how the cooling will proceed. Based on the measured O\,\textsc{vi}\,$\lambda$1038 flux, and following \cite{edgar86}, we estimate a cooling rate of 0.85 $\pm$ 0.15 (stat) $\pm$ 0.20 (sys) M$_{\odot}$ yr$^{-1}$, where the systematic uncertainty includes expectation for isobaric (1.0 $\pm$ 0.2 M$_{\odot}$ yr$^{-1}$) and isochoric (0.7 $\pm$ 0.2 M$_{\odot}$ yr$^{-1}$) cooling.

This cooling rate, which probes $\sim$10$^{5.5}$\,K gas, can be compared to both the X-ray cooling rate and the star formation rate, which probe the hot ($\sim$10$^7$\,K) and cold ($\sim$10\,K) extremes, respectively. We estimate the classical cooling rate (\.{M}$_X$ = $2L_X\mu m_p/5kT$) within the COS aperture to be $\sim$1 M$_{\odot}$ yr$^{-1}$, assuming that, on small ($<$kpc) scales, the cooling ICM is similar in morphology (clumpiness) to the UV continuum. Based on our stellar population modeling (\S3.2), we estimate an extinction-corrected star formation rate (assuming continuous star formation) within the COS aperture of 0.11 $\pm$ 0.02 M$_{\odot}$ yr$^{-1}$. This suggests that $13^{+3}_{-2}$\% of the gas cooling through 10$^{5.5}$\,K is converted to stars. 

Alternatively, some fraction of the O\,\textsc{vi} emission could come from a mixing layer, where hot electrons from the ICM penetrate the cold filaments, following \cite{fabian11} -- correcting for this would raise the inferred star formation efficiency. However, we do not find an excess of optical line emission above the expectation given the UV continuum level, as is observed in NGC~1275, suggesting that these effects are small in the filaments of Abell~1795.

For comparison, \cite{bregman06} find a total O\,\textsc{vi}-derived cooling rate of 42 $\pm$ 9 M$_{\odot}$ yr$^{-1}$ over the full core of A1795, while the extinction-corrected star formation rate \citep[assuming $\left<E(B-V)\right>$ = 0.1;][]{mcdonald12a} over the same area is only 5.2 M$_{\odot}$ yr$^{-1}$ \citep{mcdonald09}. This corresponds to an efficiency of forming stars out of the 10$^{5.5}$\,K gas of 12$^{+4}_{-2}$\%. This is also comparable to the star formation efficiency found by \cite{mcdonald11b} of 14$^{+18}_{-8}$\% based on X-ray spectroscopy and UV photometry.

\section{Implications for the Cooling Flow Problem}
Given that cooling flows are found to be $\sim$1\% efficient at converting the cooling ICM into stars \citep[e.g.,][]{odea08}, it is important to understand at what temperatures the bulk of the gas is held up. Assuming that the IMF does indeed follow \cite{salpeter55}, we find, for the filaments in Abell~1795 (where the effects of AGN feedback should be minimized), that the cooling efficiency at high temperatures is of order 100\% 
($\epsilon_{hot}$~$\equiv$~\.{M}$_{\textrm{\footnotesize O{\tiny VI}}}$/\.{M}$_X $ $\sim$ 0.7--1.0), 
while the star formation efficiency is low ($\epsilon_{cold}$~$\equiv$~SFR/\.{M}$_{\textrm{\footnotesize O{\tiny VI}}}$~$\sim$~0.11--0.16). In contrast, \cite{bregman01} found, using O\,\textsc{vi} observations from FUSE, that $\epsilon_{hot}$ was of order 10\%\ \emph{over the full cluster core} for Abell~1795, Abell~2597, and Perseus. This suggests that the cooling flow problem may be divided into two separate inefficiencies: 

\begin{itemize}
\item $\epsilon_{hot} \sim 0.1$ : globally inefficient cooling at high temperatures (10$^7$K $\rightarrow$ 10$^5$K), due to some large-scale feedback source (e.g., AGN); 
\item $\epsilon_{cold} \sim 0.1$ : locally inefficient cooling at low temperatures (10$^5$K $\rightarrow$ stars), manifesting as inefficient star formation. 
\end{itemize}

The latter, which we quantify here and in \cite{mcdonald11b}, may be low due to conduction suppressing cooling at low temperatures. For comparison, star clusters in our Galaxy have typical star formation efficiencies of $\sim$8--30\% \citep{lada03} -- in principle, star formation embedded in a hot plasma should proceed less efficiently. 

\section*{Summary}
We present deep far-UV spectroscopy of a cooling filament in Abell~1795, obtained using the \emph{Cosmic Origins Spectrograph} on HST. These data allow us to simultaneously probe the young stellar populations and the intermediate temperature (10$^{5.5}$\,K) gas. We find evidence for ongoing, in situ star formation, which suggests that the cool gas in these filaments was \emph{not} stripped from an infalling galaxy.  The detection of O\,\textsc{vi}\ emission suggests that this star formation is being fueled by condensing intracluster gas, and that the cooling is proceeding efficiently at high temperatures, contrary to what is observed on large scales in cluster cores. We propose a scenario where the two orders of magnitude disagreement between luminosity-based X-ray cooling rates and star formation in cluster cores is due to a combination of globally-inefficient cooling at high temperatures ($\epsilon_{hot} \sim 0.1$; e.g., AGN feedback) and locally-inefficient star formation at low temperatures ($\epsilon_{cold} \sim 0.1$).

\section*{Acknowledgements} 
M.\,M. acknowledges support provided by NASA through a Hubble Fellowship grant from STScI and through HST GO-12992 contract NAS5-26555. S.\,V. acknowledges support from a Senior NPP Award held at NASA-GSFC. S.\,E. acknowledges support from SAO subcontract SV2-82023 under NASA contract NAS8-03060.


\end{document}